\documentstyle[12pt]{article}
\input epsf.sty
\newcommand{\be}{\begin{equation}}
\newcommand{\ee}{\end{equation}}
\newcommand{\bea}{\begin{eqnarray}}
\newcommand{\eea}{\end{eqnarray}}
\newcommand{\al}{\alpha}
\newcommand{\bt}{\beta}

\newcommand{\zt}{\zeta}

\newcommand{\kp}{\kappa}
\newcommand{\lm}{\lambda}
\newcommand{\Lm}{\Lambda}

\newcommand{\Om}{\Omega}

\newcommand{\nn}{\nonumber}

\begin{document}
\title{Brane Cosmology from Heterotic String Theory}
\author{Apostolos Kuiroukidis\thanks{E-mail:
kouirouki@astro.auth.gr}\\
and\\
Demetrios B. Papadopoulos\thanks{E-mail:
papadop@astro.auth.gr}\\
\small Department of Physics, A.U.Th., 541 24 Thessaloniki, GREECE\\
}

\maketitle
\begin{abstract}
We consider brane cosmologies within the context of
five-dimensional effective actions with ${\cal O}(\al ^{'})$
higher curvature corrections. The actions are compatible with
bulk string amplitute calculations from heterotic string theory.
We find wrapped solutions that satisfy the field equations in
an {\it approximate} but acceptable manner given their complexity,
where
the internal, four-dimensional, scale factor is naturally inflating,
having an exponential De-Sitter form. The temporal dependence of
the metric components is non-trivial so that this metric cannot
be factored as in a conformally flat case. The effective Planck
mass is finite and the brane solutions can localize
four-dimensional gravity while the four-dimensional gravitational
constant varies with time. The Hubble constant can be
freely specified through the initial value of the scalar field,
to conform with recent data.\\

\end{abstract}

\newpage

\section{Introduction}\
Recent developments in string theory suggest that matter and
gauge interactions may be confined on a brane, embedded in a
higher dimensional space (bulk), while gravity can propagate
into the bulk (for reviews see \cite{epap}, \cite{maar1}).
Within this context several toy models
have been constructed to address such issues as the hierarchy
and cosmological constant problems \cite{aluk}.
In particular, the large hierarchy between the Standard Model and
Planck scales could be explained for an observer on a negative
tension flat brane, if the extra dimension was taken to be compact
\cite{rand1}. The possibility of a large non-compact dimension
was realized in \cite{nark}, while it was shown in \cite{rand2}
that warping of five-dimensional space could lead to localization
of gravity on the brane, even though the size of the extra dimension
was of infinite proper length.

A simple, interesting alternative model
has been considered in \cite{arka}, \cite{kach},
where a bulk scalar field $\phi $ is coupled to
the brane tension $T_{br}$. This is the all-loop contribution to the
vacuum energy density of the brane, from the Standard Model fields.
For the 4D cosmological constant problem considered there, solutions
of the field equations were found, which localize gravity, but possess
naked singularities at finite proper distance. This proper distance
is given by $y_{c}=1/\kp _{(5)}^{2}T_{br}$ where the five and
four-dimensional Planck scales $k_{(5)}^{2}=M_{(5)}^{-3}$,
$\kp _{(4)}^{2}=M_{(4)}^{-2}$ are related by
\be
T_{br}=\frac{\kp _{(4)}^{2}}{\kp _{(5)}^{4}}=
\frac{M_{(5)}^{6}}{M_{(4)}^{2}}.
\ee
Then if we momentarily identify the 4D cosmological term with the
brane tension \\
$\lm =T_{br}\sim (1TeV)^{4}=(10^{3}GeV)^{4}\sim 10^{-64}
M_{(4)}^{4}$ we obtain
\be
M_{(5)}\simeq 10^{8}GeV,\; \; y_{c}\simeq 1mm,
\ee
which is acceptable by present day experiments.

It was later realized that the bulk action should also contain
the Gauss-Bonnet (GB) term
\be
{\cal L}_{GB}=R^{2}-4R_{ab}R^{ab}+R_{abcd}R^{abcd},
\ee
which is the leading quantum gravity correction,
and the only to provide second order field equations.
Some of the early works on the GB gravity include
\cite{egb}, while the corresponding brane cosmology has been studied,
among others, in \cite{rgb}. The corresponding generalized junction
conditions appeared in \cite{jun}.

If we consider the action \cite{lidsey}
\bea
\label{lids}
S&=&\int d^{5}x\sqrt{|g|}\left[R-\frac{1}{2}(\nabla \phi )^{2}\right.
-\frac{1}{2}e^{-\phi }(\nabla \sigma ){^2}
\left.-\Lambda e^{-2\phi }\right]\nn \\
&+&\sum _{i=1}^{2}(-1)^{i}\sqrt{24}\Lambda \int d^{4}x
\sqrt{|g_{i}|}e^{-\phi }
\eea
then Eq. (\ref{lids}) introduces
the fact that the bulk cosmological constant couples to the
scalar field through the exponential potential term.\\
However serious arguments
were given in \cite{mavr1}-\cite{binet}
that one must consider a
modified action, instead of the usual one used in [10]
where the GB term is multiplied by a constant.
The modified action comes from heterotic string amplitude
calculations, truncated to five dimensions \cite{mets},
\cite{gross}. In this action the GB combination enters
through a multiple of the exponential of the dilaton field.\\

To write this action we use the
constants $\kappa _{(i)}^{2}=8\pi G_{i}=M_{(i)}^{2-i},\; i=4,5$,
which represent the fundamental, five-dimensional and the effective
four-dimensional Planck masses. The bulk cosmological constant
has dimensions $\Lambda _{(5)}=[energy]^{2}$, defining an inverse
length scale squared, effectively an $AdS_{5}$ curvature.
Also a brane cosmological constant would have dimensions
$\Lambda _{(4)}=[energy]^{2}$ and divided both by the respective
mass scales $\kp _{(i)}^{2}$ would have dimensions similar to
the corresponding brane tensions.
With this normalization, the action used
in \cite{mavr1}-\cite{binet} is dimensionless,
and can be written as
\bea
\label{no1}
S_{1}=\frac{1}{2\kappa _{(5)}^{2}}\int d^{5}x\sqrt{|g_{(5)}|}
\left\{R-\zeta (\nabla \phi )^{2}+\alpha \right.
e^{-\zeta \phi }[R_{GB}^{2}+c_{2}\zeta ^{2}
\left.(\nabla \phi )^{4}]\right\}
\eea
Here $\alpha =\alpha ^{'}/8g_{s}^{2}$, with $l_{s}=\sqrt{\alpha ^{'}}$
the string length, and $g_{s}=exp(-\phi _{0})$ the string coupling constant,
where $\phi _{0}$ is the vacuum expectation value of the dilaton
field. Also $\zeta =(4/3)$ and $c_{2}=(D-4/D-2)$ and in our case
$c_{2}=(1/3)$.\\
In addition we take as
\bea
\label{no2}
S_{2}=\frac{-1}{2\kappa _{(5)}^{2}}\int d^{5}x\sqrt{|g_{(5)}|}
\left[2\Lambda _{(5)}V(\phi )\right]
\eea
Using this normalization the bulk potential and also
$\phi $ is dimensionless. In \cite{binet}, it was taken
as $V(\phi )=e^{\zt \phi }$ and this will also be
our choice.\\
So with this normalization the various $S_{i}$ are
dimensionless. We take as
\be
S=S_{1}+S_{2}
\ee
Solutions to the first part of the
action were studied in \cite{mavr1}-\cite{char1}.
However, due to the severe complexity of
the equations of motion only metrics of the form
\be
ds^{2}=e^{2A(y)}\eta _{\mu \nu}dx^{\mu }dx^{\nu }+dy^{2}
\ee
were considered. This choice limits the range of solutions for
realistic internal, four-dimensional spacetimes. One of the
interesting cases to consider is an inflating internal space.
This is because it is generally accepted that the Universe now
undergoes accelerated expansion \cite{acc}.
In this paper we write the field equations for the action
of Eq. (7), and use a metric with internal De-Sitter-like scale
factor. The extreme complexity of the field equations does not
allow for an exact, mathematical solution. However for a
continuous range of one parameter of the model, three of the
field equations coincide, with acceptable accuracy, permiting
a class of models with interesting characteristics to appear.
First the action is a realistic one, coming from string theory.
Also the brane, located at $y=0$, has an inflating internal scale
factor. Gravity is localized on the brane, so the effective
four-dimensional Planck mass is finite. The class of models
does not obey a fine-tuning condition, in the sense that the
bulk cosmological constant only controls the temporal evolution
of the scalar field. The Hubble constant is determined from the
initial value of the scalar field so that it can be adjusted to
any value without fine-tuning of the parameters of the action.

\section{Equations of Motion}
The metric will be of the form
\be
\label{metr}
ds^{2}=-n^{2}(t,y)dt^{2}+a^{2}(t,y)h_{ij}(x)dx^{î}dx^{j}+
b^{2}(t,y)dy^{2}
\ee
The three-metric from Eq. (\ref{metr})
is assumed to represent a maximally symmetric space
\be
^{(3)}ds^{2}=a^{2}h_{ij}(x)dx^{i}dx^{j}=a^{2}
\left[\frac{dr^{2}}{1-kr^{2}}+r^{2}d\Om _{II}^{2}\right]
\ee
with scalar three-curvature $^{(3)}R=6k/a^{2},\; \; k=0,\pm 1$.\\
Variation of the action with respect to the five
dimensional metric gives:
\bea
\label{genaction}
G_{\mu \nu }
&-&\zt (\nabla _{\mu }\phi )(\nabla _{\nu }\phi )
+\frac{\zt }{2}g_{\mu \nu }(\nabla \phi )^{2}+g_{\mu \nu }\Lm _{(5)}V(\phi )
+2\al e^{-\zt \phi }H_{\mu \nu }+\nn \\
&+&4\al P_{\mu \al \nu \bt }
\nabla ^{\al }\nabla ^{\bt }(e^{-\zt \phi })+
\al e^{-\zt \phi }c_{2}\zt ^{2}
[2(\nabla \phi )^{2}\nabla _{\mu }\phi \nabla _{\nu }\phi
-\frac{1}{2}g_{\mu \nu }(\nabla \phi )^{4}]=0\nn \\
\eea
Here we have (\cite{binet}, \cite{der})
\bea
\label{lov}
H_{\mu \nu }&=&RR_{\mu \nu }-2R_{\mu \al }R^{\al }_{\nu }-2R^{\al \bt }
R_{\mu \al \nu \bt }+R_{\mu }^{abc}R_{\nu abc}-\nn  \\&-&\frac{1}{4}g_{\mu \nu }
R_{GB}^{2}\nn \\
P_{\mu \al \nu \bt }&=&R_{\mu \al \nu \bt }+R_{\mu \bt }g_{\nu \al }
+R_{\nu \al }g_{\mu \bt }-R_{\mu \nu }g_{\al \bt }-R_{\al \bt }g_{\mu \nu }
+\nn \\
&+&\frac{1}{2}R(g_{\mu \nu }g_{\al \bt }-g_{\mu \bt }g_{\nu \al })
\eea
Varying with respect to $\phi $ we obtain
\bea
\label{fifi}
2\zt \nabla ^{2}\phi
-\al \zt e^{-\zt \phi }
[R_{GB}^{2}-3c_{2}\zt ^{2}(\nabla \phi )^{4}]-
2\Lm _{(5)}V^{'}(\phi )-
\nn \\-4\al c_{2}\zt ^{2}e^{-\zt \phi }
[(\nabla \phi )^{2}\nabla ^{2}\phi +2\nabla ^{\mu }\phi
\nabla ^{\nu }\phi \nabla _{\mu }\nabla _{\nu }\phi ]=0
\eea
Greek indices
denote five-dimensional components (0,1,2,3;5), while Latin
three-dimensional.
\newpage
The (00)-component of the generalized Einstein's equations is
\bea
\label{zz}
3n^{2}F+\frac{3n^{2}}{b^{2}}
\left(-\frac{a^{''}}{a}+\frac{a^{'}b^{'}}{ab}+\frac{\dot{a}b\dot{b}}{an^{2}}\right)
-\frac{2}{3}(\dot{\phi })^{2}
-\frac{2}{3}\frac{n^{2}}{b^{2}}(\phi ^{'})^{2}
-n^{2}\Lm _{(5)}V(\phi )+\nn \\
+2\al e^{-4\phi /3}H_{00}
+\frac{16\al }{3b^{4}}P_{0505}
\left[-\phi ^{''}+\frac{4}{3}(\phi ^{'})^{2}+\frac{b\dot{b}}{n^{2}}\dot{\phi }\right.
\left.+\frac{b^{'}}{b}\phi ^{'}\right]e^{-4\phi /3}+\nn \\
+\frac{16\al }{3a^{4}}P_{0i0j}h^{ij}
\left(\frac{a\dot{a}}{n^{2}}\dot{\phi }-\frac{aa^{'}}{b^{2}}\phi ^{'}\right)
e^{-4\phi /3}
+\frac{32\al }{27}e^{-4\phi /3}(\nabla \phi )^{2}
\left[\frac{3}{4}\dot{\phi }^{2}+\right.
\left.\frac{1}{4}\frac{n^{2}}{b^{2}}(\phi ^{'})^{2}\right]=0
\eea
where we use the conventions
\be
F:=\frac{1}{a^{2}}
\left(\frac{\dot{a}^{2}}{n^{2}}-\frac{(a^{'})^{2}}{b^{2}}\right)
+\frac{k}{a^{2}},\; \; \;
(\nabla \phi )^{2}=-\frac{1}{n^{2}}\dot{\phi }^{2}+\frac{1}{b^{2}}
(\phi ^{'})^{2}
\ee
A dot denotes a partial derivative
with respect to the time, while a prime denotes derivative
with respect to the extra dimension, denoted by y.\\
The (05)-component of the Einstein's equations is
given by
\bea
\label{zf}
-3\left(\frac{\dot{a}^{'}}{a}-\frac{\dot{a}n^{'}}{an}-\frac{a^{'}\dot{b}}{ab}\right)
-\frac{4}{3}\dot{\phi }\phi ^{'}+2\al e^{-4\phi /3}H_{05}
+\frac{32\al }{27}e^{-4\phi /3}(\nabla \phi )^{2}\dot{\phi }\phi ^{'}+\nn \\
+\frac{16\al }{3b^{2}n^{2}}P_{0505}
\left[-\dot{\phi }^{'}+\frac{4}{3}\dot{\phi }\phi ^{'}\right.
\left.+\frac{n^{'}}{n}\dot{\phi }+\frac{\dot{b}}{b}\phi ^{'}\right]e^{-4\phi /3}
+\frac{16\al }{3a^{4}}P_{0i5j}h^{ij}
\left(\frac{a\dot{a}}{n^{2}}\dot{\phi }-\frac{aa^{'}}{b^{2}}\phi ^{'}\right)
e^{-4\phi /3}=0\nn \\
\eea
The (55)-component is given by
\bea
\label{ff}
-3b^{2}F+\frac{3b^{2}}{n^{2}}
\left(-\frac{\ddot{a}}{a}+\frac{\dot{a}\dot{n}}{an}+\frac{a^{'}nn^{'}}{ab^{2}}\right)
-\frac{2}{3}(\phi ^{'})^{2}-\frac{2}{3}
\frac{b^{2}}{n^{2}}(\dot{\phi })^{2}+b^{2}\Lm _{(5)}V(\phi )+\nn \\
+2\al e^{-4\phi /3}H_{55}
+\frac{16\al }{3n^{4}}P_{0505}
\left[-\ddot{\phi }+\frac{4}{3}(\dot{\phi })^{2}+\frac{\dot{n}}{n}\dot{\phi }\right.
\left.+\frac{nn^{'}}{b^{2}}\phi ^{'}\right]e^{-4\phi /3}+\nn \\
+\frac{16\al }{3a^{4}}P_{5i5j}h^{ij}
\left(\frac{a\dot{a}}{n^{2}}\dot{\phi }-\frac{aa^{'}}{b^{2}}\phi ^{'}\right)
e^{-4\phi /3}
+\frac{32\al }{27}e^{-4\phi /3}(\nabla \phi )^{2}
\left[\frac{3}{4}(\phi ^{'})^{2}+\frac{1}{4}\frac{b^{2}}{n^{2}}(\dot{\phi })^{2}\right]
=0\nn \\
\eea
\newpage
The (ij)-component of the generalized Einstein's equations
is a multiple of $h_{ij}$. Setting this proportionality term
equal to zero gives
\bea
-2(\frac{a\ddot{a}}{n^{2}}-\frac{a\dot{a}\dot{n}}{n^{3}}
-\frac{aa^{'}n^{'}}{nb^{2}})-2
(-\frac{aa^{''}}{b^{2}}+\frac{aa^{'}b^{'}}{b^{3}}+\frac{a\dot{a}\dot{b}}{bn^{2}})
-a^{2}F-\nn \\-
\frac{a^{2}}{n^{2}}
(\frac{\ddot{b}}{b}-\frac{nn^{''}}{b^{2}}+\frac{nn^{'}b^{'}}{b^{3}}-
\frac{\dot{n}\dot{b}}{nb})+\frac{2}{3}a^{2}(\nabla \phi )^{2}+a^{2}\Lm _{(5)}
V(\phi )+
2\al e^{-4\phi /3}(\frac{1}{3}h^{ij}H_{ij})-
\nn \\-\frac{8\al }{27}
e^{-4\phi /3}a^{2}(\nabla \phi )^{4}+\frac{16\al }{9n^{4}}P_{i0j0}h^{ij}
[-\ddot{\phi }+\frac{4}{3}(\dot{\phi })^{2}+\frac{\dot{n}}{n}\dot{\phi }
+\frac{nn^{'}}{b^{2}}\phi ^{'}]e^{-4\phi /3}-\nn \\
-\frac{32\al }{9n^{2}b^{2}}P_{0i5j}h^{ij}
[-\dot{\phi }^{'}+\frac{4}{3}\dot{\phi }\phi ^{'}+\frac{n^{'}}{n}\dot{\phi }
+\frac{\dot{b}}{b}\phi ^{'}]e^{-4\phi /3}+\nn \\
+\frac{16\al }{9b^{4}}P_{i5j5}h^{ij}
[-\phi ^{''}+\frac{4}{3}(\phi ^{'})^{2}+\frac{b\dot{b}}{n^{2}}\dot{\phi }
+\frac{b^{'}}{b}\phi ^{'}]e^{-4\phi /3}+\nn \\
+\frac{16\al }{9a^{4}}P_{ikjm}h^{ij}h^{km}
(\frac{a\dot{a}}{n^{2}}\dot{\phi }-\frac{aa^{'}}{b^{2}}\phi ^{'})
e^{-4\phi /3}=0\nn \\
\eea
The contraction $h^{ij}H_{ij}$ is given in Appendix I.

\section{Reduction and an Exact Fine-Tuned Solution}\
We consider flat spatial sections ($k=0$)
in Eq. (10) and introduce the following ansatz:
\bea
a=a(t)A(y),\; n=n(t)N(y),\; b=b(t)B(y), \nn \\
\phi =\sigma (t)+\phi (y)
\eea
where
\bea
a(t)=a_{0}e^{Ht},\; b(t)=b_{0}e^{-2\sigma _{1}Ht/3},\; \nn \\
n(t)=n_{1}b(t),\; \; \sigma (t)=\sigma _{1}Ht+\sigma _{2}
\eea
with $H,a_{0},b_{0},\sigma _{1},n_{1},\sigma _{2}$ constants.
The field equations become exactly those presented in Appendix I.
Now we make the assumption
\bea
A(y)=A_{0}e^{f_{1}y},\; \; B(y)=B_{0}e^{f_{2}y}\nn \\
N(y)=N_{0}e^{f_{2}y},\; \; \phi (y)=-\frac{3}{2}f_{2}y+\phi _{0}
\eea
We obtain five equations constraining the numerical
parameters of our solution. Inspecting the metric and the
resulting equations it is easy to see that we can set
$a_{0}=1=A_{0}$. It appears, therefore, that there exist eight (8)
independent parameters, namely
$H,\; f_{1},\; f_{2},\; N_{1}:=n_{1}N_{0},\; ,
\Phi _{0}:=\sigma _{2}+\phi _{0},\; \sigma _{1}, b_{0}$
and $B_{0}$. \\
The solution for the metric is therefore written
\bea
ds^{2}=&-&b_{0}^{2}N_{1}^{2}e^{-4\sigma _{1}Ht/3}e^{2f_{2}y}dt^{2}
+e^{2Ht}e^{2f_{1}y}[dx_{1}^{2}+dx_{2}^{2}+dx_{3}^{2}]+\nn \\
&+&b_{0}^{2}B_{0}^{2}e^{-4\sigma _{1}Ht/3}e^{2f_{2}y}dy^{2}
\eea
However a closer inspection shows that rescaling the metric
and redefining the internal coordinates $x_{j}$,
we can set $b_{0}=1=N_{1}=B_{0}$. This also occurs from the five
field equations. So there exist five (5) independent
constants $H,\; f_{1},\; f_{2},\; \Phi _{0},\; \sigma _{1}$
and the metric takes the final form
\bea
ds^{2}=&-&e^{-4\sigma _{1}Ht/3}e^{2f_{2}y}dt^{2}+
e^{2Ht}e^{2f_{1}y}[dx_{1}^{2}+dx_{2}^{2}+dx_{3}^{2}]+\nn \\
&+&e^{-4\sigma _{1}Ht/3}e^{2f_{2}y}dy^{2}
\eea
The solution for the scalar field is
\be
\phi (t,y)=\sigma _{1}Ht-\frac{3}{2}f_{2}y+\Phi _{0}
\ee
Inspecting the field equations we see that the following
\bea
\sigma _{1}=0,\; \; f_{1}=f_{2}:=f,\; \; H^{2}=f^{2},\nn \\
e^{4\Phi _{0}/3}=2\al f^{2},\; \; f^{2}=-\frac{4}{3}\Lm _{(5)},
\nn \\\frac{8}{3}\al \Lm _{(5)}=-1
\eea
is an {\it exact} fine-tuned solution, which means that the
cosmological constant and the Hubble constant are directly
related. This solution though exact does not localize
four-dimensional gravity on the brane, located at $y=0$.
\section{Classes of Non Fine-Tuned Solutions}\
We set $f_{1}=f_{2}=f$ into the five equations of Appendix I
and use the metric, Eq. (23). We thus have four constants,
namely
$f,\; H,\; \Phi _{0},\; \sigma _{1}$.
From the (05)-equation we obtain
\be
f^{2}=\frac{2}{9}
\left(1-\frac{8}{3}\sigma _{1}-\frac{2}{9}\sigma _{1}^{2}\right)
H^{2}
\ee
Eq. (26) constrains the allowed value of $\sigma _{1}$ so
that $-12.36\simeq -6-\frac{9\sqrt{2}}{2}\leq\sigma _{1}\leq
-6+\frac{9\sqrt{2}}{2}\simeq 0.364$

We consider now that $|\sigma _{1}|$ is much smaller
than unity. Then we keep the first order terms in
the field equations of Appendix I, with respect
to $\sigma _{1}$. We obtain
\be
f^{2}=\frac{2}{9}
\left(1-\frac{8}{3}\sigma _{1}\right)
H^{2}
\ee
The last term, in the parentheses of Eq. (26), contributes
0.02 to the sum only, validating our approximation,
for the above range of $\sigma _{1}$.
Using Eq. (27) into the $\phi -$equation, we
obtain for the bulk cosmological constant,
\bea
\Lm _{(5)}e^{4\Phi _{0}/3}=&-&(1+\frac{1}{3}\sigma _{1})H^{2}
+\nn \\&+&\al e^{-4\Phi _{0}/3}
\left(-\frac{62}{9}+\frac{20}{27}\sigma _{1}\right)
H^{4}
\eea
From the addition of the (00) and (55) components,
we obtain
\be
(4\sigma _{1}+3)=2\al e^{-4\Phi_{0}/3}
(\frac{4}{3}-\frac{16}{9}\sigma _{1})H^{2}
\ee
From the (00)-equation, using Eq. (28),
we get
\bea
(\sigma _{1}+3)+\al e^{-4\Phi _{0}/3}
[-\frac{4}{3}+\frac{28}{9}\sigma _{1}]H^{2}=0
\eea
Finally the (ij)-equation,
with the aid of Eq. (28), gives
\bea
3(\sigma _{1}+1)+\frac{4}{81}\al e^{-4\Phi _{0}/3}
[\frac{503}{3}\sigma _{1}+85]H^{2}=0
\eea
Given the complexity of the field equations, the
simplicity of the above, reduced field equations,
is quite interesting.

\begin{figure}[h!]
\centerline{\mbox {\epsfbox{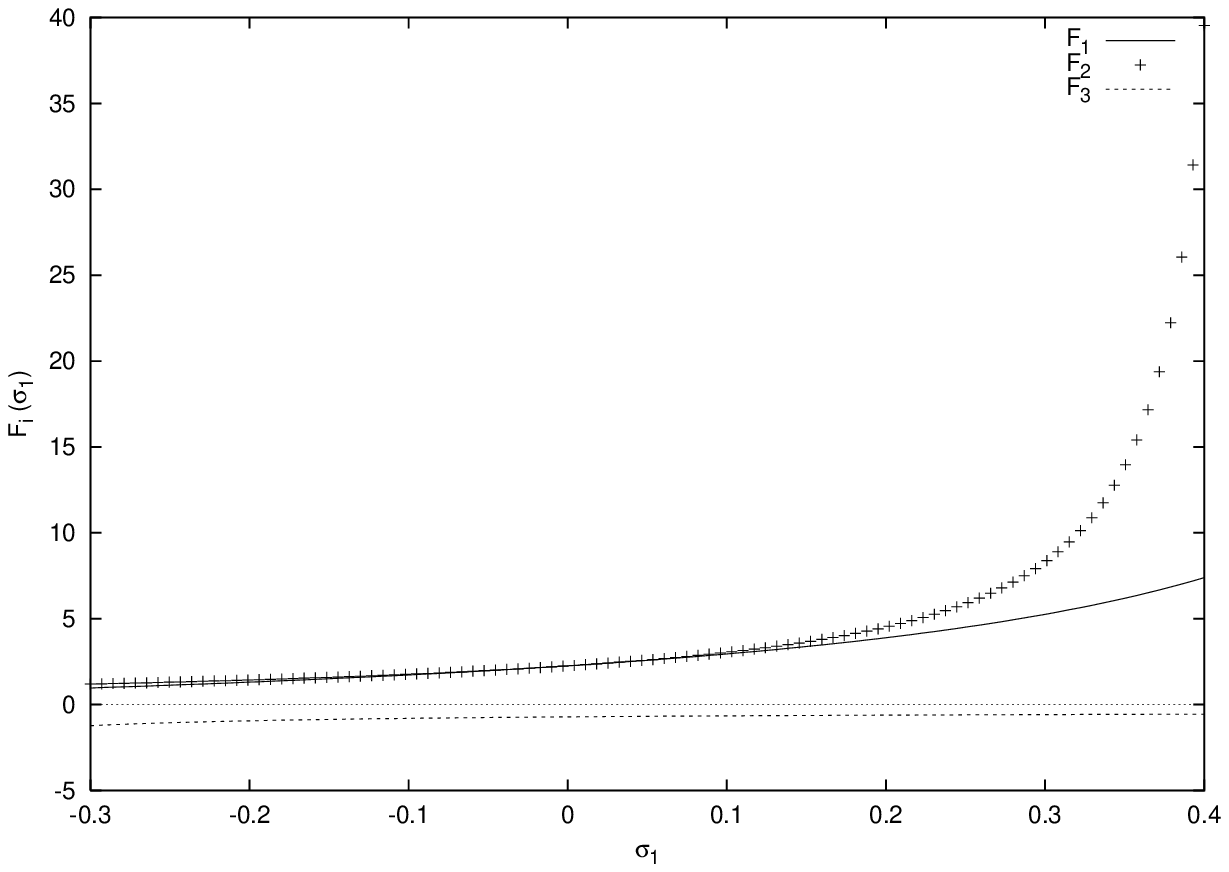}}}
\caption{The functions $F_{i}(\sigma _{1})$ over
$-0.3\leq \sigma _{1}\leq 0.4$}
\end{figure}

\newpage

\begin{figure}[h!]
\centerline{\mbox {\epsfbox{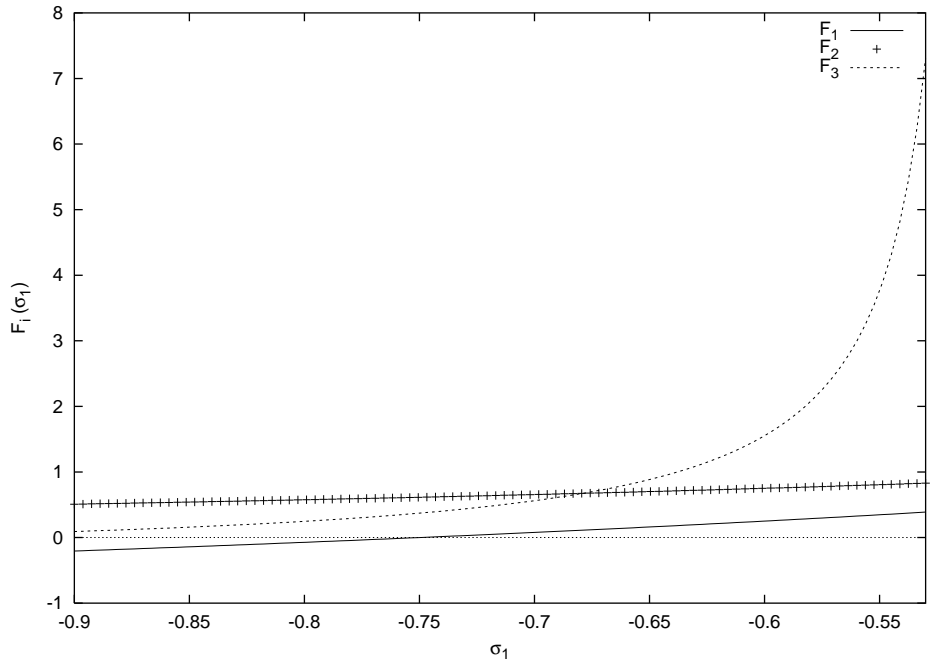}}}
\caption{The functions $F_{i}(\sigma _{1})$ over
$-0.9\leq \sigma _{1}\leq -0.5$}
\end{figure}

We define $F(\sigma _{1}):=\al e^{-4\Phi _{0}/3}H^{2}$.
This function assumes the forms $F_{1},\; F_{2},\; F_{3}$ as these
emerge from Eqs. (29), (30) and Eq. (31), respectively.

Given the extreme complexity of the field equations it is
surprising that over the whole range of $-0.3\leq \sigma _{1}\leq 0.2$
the two functions $F_{1}(\sigma _{1}),\; F_{2}(\sigma _{2})$ coincide.
Also in this interval and in $-0.9\leq \sigma _{1}\leq -0.55$ these
three functions coincide with adequate accuracy as is shown in
Figs. (1)-(2). The difference in their value is suppressed in this
intervals, as compared to other intervals.

Our aim here is not to give mathematically exact solutions, but to
stress that with acceptable accuracy we can find cosmological models
that show many interesting features. We assume therefore,
given also the approximation we have made
for small $|\sigma _{1}|$, that three of the five field equations,
namely the (00)+(55), (00) and (ij) components, give
\be
F(\sigma _{1}):=\al e^{-4\Phi _{0}/3}H^{2}\simeq
F_{1}(\sigma _{1})=\frac{3+4\sigma _{1}}
{\left(\frac{4}{3}-\frac{16}{9}\sigma _{1}\right)}
\ee
where $\sigma _{1}\in (-0.9,-0.55)\cup (-0.3,0.2)$.

Therefore the Hubble constant in string units
is given in terms of the initial value of the scalar field as
\be
\al H^{2}=e^{4\Phi _{0}/3}\frac{3+4\sigma _{1}}
{\left(\frac{4}{3}-\frac{16}{9}\sigma _{1}\right)}
\ee
and so $f^{2}$, which determines the spatial dependence of the
scalar field through Eq. (24), is determined through Eq. (27).\\
Multiplying Eq. (28) by $\al $, and using Eq. (33),
we obtain the bulk cosmological constant in string units,
\be
\al \Lm _{(5)}=\frac{(3+4\sigma _{1})}
{\left(\frac{4}{3}-\frac{16}{9}\sigma _{1}\right)^{2}}
\left[-22-24\sigma _{1}+\frac{32}{9}\sigma _{1}^{2}\right]
\ee
The dependence of the bulk comological constant on
$\sigma _{1}$ is shown in Fig. (3).

\begin{figure}[h!]
\centerline{\mbox {\epsfbox{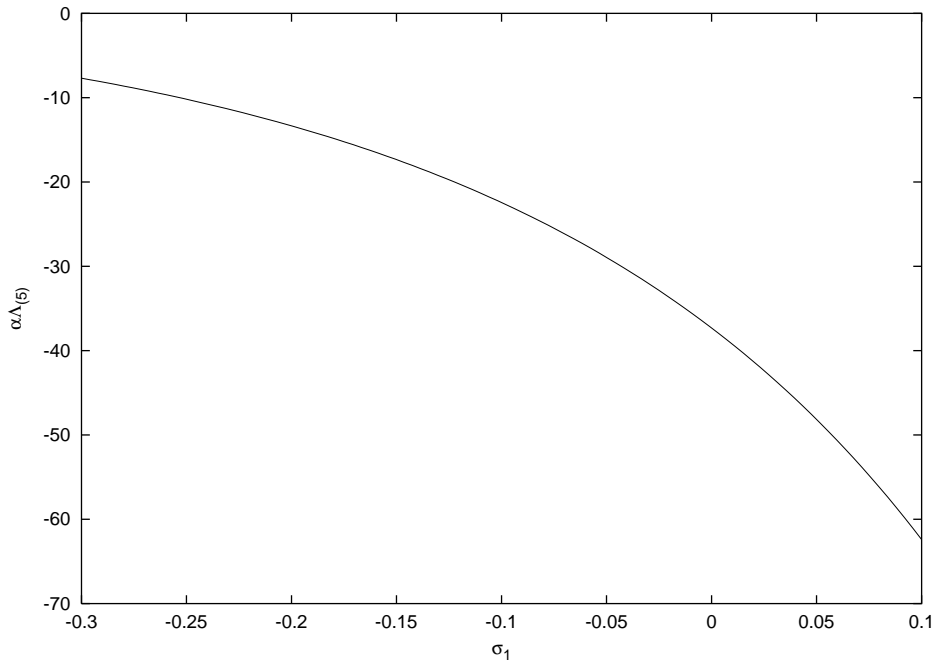}}}
\caption{The bulk cosmological constant over
$-0.3\leq \sigma _{1}\leq 0.1$}
\end{figure}

Thus we do not have to fine-tune the Hubble
constant or the bulk cosmological constant $\Lm _{(5)}$. These, as
well the constant $f$, are specified in terms of $\sigma _{1}$,
which controls the temporal evolution of the scalar field through
Eq. (24), and the initial value $\Phi _{0}$. So $\sigma _{1}$
determines the bulk cosmological constant through Eq. (34), but
the Hubble constant $H$ is determined by the initial value of the
scalar field as well, (Eq. (33)) and can be freely adjusted. The
metric is given by Eq. (23) (with $f_{1}=f_{2}=f$) and the scalar
field by Eq. (24).

Finally if we assume that $f_{1}\neq f_{2}$,
there will, in general, exist {\it exact} solutions of the five
field equations, for the five independent constants
$f_{1}, f_{2}, H, \sigma _{1}, \Phi _{0}$. These will also
be determined in terms of $\Lm _{(5)}$. In order to determine them
one has to resort to numerical methods and this will be the subject
of a future work.

\section{Localization of Gravity}\
The four dimensional scalar curvature, with flat spatial sections
for the metric of\\ Eq. (10), is given by  $^{(4)}R=
(\frac{6}{n^{2}})(\frac{\ddot{a}}{a}-\frac{\dot{a}\dot{n}}{an}
+(\frac{\dot{a}}{a})^{2})$. If one substitutes the five-dimensional
scalar curvature $R$ and the Gauss-Bonnet contribution
into the action functional, then the integrated coefficient
of the four-dimensional scalar curvature will give the effective
four-dimensional Planck mass as perceived by a four-dimensional
observer, located on the brane \cite{mavr1}-\cite{binet}.
Doing this in a careful manner we obtain
\bea
M_{Pl}^{2}&=&M_{s}^{3}\int_{0}^{R} dy b
\left[1+\frac{8\al e^{-4\phi /3}}{b^{2}}\left(-\frac{a^{''}}{a}+\right.\right.
\left.\left.\frac{a^{'}b^{'}}{ab}+\frac{\dot{a}b\dot{b}}{an^{2}}\right)\right]
=\nn \\
&=&M_{s}^{3}e^{-2\sigma _{1}Ht/3}\frac{1}{|f|}
[1-e^{-|f|R}]
\left(1-\frac{16}{3}\sigma _{1}\frac{3+4\sigma _{1}}{\frac{4}{3}-\frac{16}{9}\sigma _{1}}\right)
\eea
This must be finite as $R\rightarrow +\infty $, so this is why
we have chosen the $f<0$ solution of Eq. (27). For the negative
y-direction we can choose $f>0$ and match the two bulk solutions
continuously on the brane.

The quantity in the large parentheses of Eq. (35) is positive
for the range of \\
$\sigma _{1}\in (-0.9,-0.55)\cup (-0.3,0.1)$, giving an overall
positive Planck mass. So these cosmological
solutions can localize four-dimensional gravity. Also we observe
that we naturally obtain a time-varying gravitational constant.
This occurs due to the non-trivial, different time dependence
of the various metric components of the metric, Eq. (23), and also
due to the fact that we have the exponential $e^{4\Phi _{0}/3}$
in Eq. (33). The last comes from the coefficient of the GB
term of the action, Eq. (5).

\section{Discussion}\
We have considered the action that results from a consistent
truncation to five dimensions of the heterotic string. This
action has higher order gravitational corrections of the
form of the Gauss-Bonnet term. However this term enters the
expansion, multiplied not only by the string constant but also
by the exponential of the scalar field. This makes the
resulting field equations very complicated, compared to the
usual case when the exponential term is absent. Due to this
fact, only solutions with the four-dimensional Poincare-invariant
form of Eq. (8), have been considered in the literature.
In this paper we showed that there exist one exact
and a family of approximate solutions, continuously dependent on
the parameter $\sigma _{1}$, with the metric given by Eq. (23).

The important features of these cosmological models can be
summarized as follows: The metric cannot be factored as in
a simple, conformally flat case, since in general
$\sigma _{1}\neq -(3/2)$. The temporal and spatial dependence
of the metric components is non-trivial and does not allow a
conformally flat solution even in the four-dimensional subcase.
The metric cannot be simplified any further by a coordinate
transformation.\\
Also the brane, situated at $y=0$, can localize four-dimensional
gravity. This is due to the fact that
the four-dimensional effective Planck mass is finite and positive
for the allowed range of the parameter $\sigma _{1}$.\\
More importantly the parameters of the theory need not be fine-tuned.
By this we mean that the parameters of our action such as the bulk
cosmological constant need not be fine-tuned to a specific value
in order to obtain a desired solution. The bulk cosmological
constant, is directly related to the temporal evolution of the
scalar field, i.e., to $\sigma _{1}$. The Hubble constant however is
freely determined from the initial value of the scalar field and so
a proper choice of the last can accomodate observational data.\\
Finally the four-dimensional gravitational constant varies with time
and follows the exponential expansion of the four-dimensional scale
factor [5], [21].

The action used in this paper is a realistic one because it
occurs in a consistent way from heterotic string theory [17]. So
it is important to search for brane cosmological solutions that
give realistic four-dimensional cosmological models. Because
the field equations are complicated, the use of combined numerical
and analytical techniques is necessary. One can numerically
search for solutions with the metric assuming the form of Eq. (23)
and without any other assumption. Work along these lines is
in progress.
\newpage

\section{Appendix I}\
Contracting the first of Eqs. (\ref{lov}) we get
\bea
\label{contr}
h^{ij}H_{ij}&=&-\frac{1}{4}a^{2}R_{GB}^{2}+\frac{a^{2}}{n^{2}}H_{00}
-\frac{a^{2}}{b^{2}}H_{55}
\eea

\section*{Acknowledgments}\
One of the authors (A.K)
would like to thank the Greek State Scholarships Foundation
(I.K.Y) for postdoctoral financial support,
under contract No. 422, during this work.

\end{document}